\documentclass[amsmath,amssymb,aps,prl,reprint,tightenlines]{revtex4-2}

\usepackage{graphicx}
\usepackage{float}
\usepackage{dcolumn}
\usepackage{bm}
\usepackage{xcolor}
\usepackage{natbib}
\usepackage{setspace}
\usepackage{amsmath}              
\usepackage{amssymb}              
\usepackage{amsfonts}             
\usepackage[amssymb]{SIunits}     
\usepackage[applemac]{inputenc}
\usepackage[english]{babel}
\newcommand{\bsfA}{\textbf{\textsf A}}
\newcommand{\bsfP}{\textbf{\textsf P}}
\newcommand{\bsfL}{\textbf{\textsf L}}

\newcommand{\rev}[1]{\textcolor{black}{#1}}

\begin{document}

\title{Rheology of suspensions of flat elastic particles}

\author{Jens Eggers, Tanniemola B. Liverpool, and Alexander Mietke}
\affiliation{School of Mathematics, University of Bristol, Fry Building, Bristol BS8 1UG, United Kingdom}

\begin{abstract}
We consider a suspension of non-interacting flat elastic particles in a Newtonian fluid. We model a flat shape as three beads, carried along by the flow according to Stokes' law, and connected by nonlinear springs, chosen such that the energy is quadratic in the area. In analogy with common dumbbell models involving two beads connected by linear springs, we solve the stochastic equations of motion exactly to compute the constitutive law for the stress tensor of a flat elastic particle suspension. A lower convected time derivative naturally arises as part of the constitutive law, but surprisingly the rheological response in strong extensional and strong contracting flows is similar to that of the classical Oldroyd-B model associated with dumbbell suspensions. 
\end{abstract}

\maketitle

\begin{figure}[!t]
	\centering	
\includegraphics[width=0.48\textwidth]{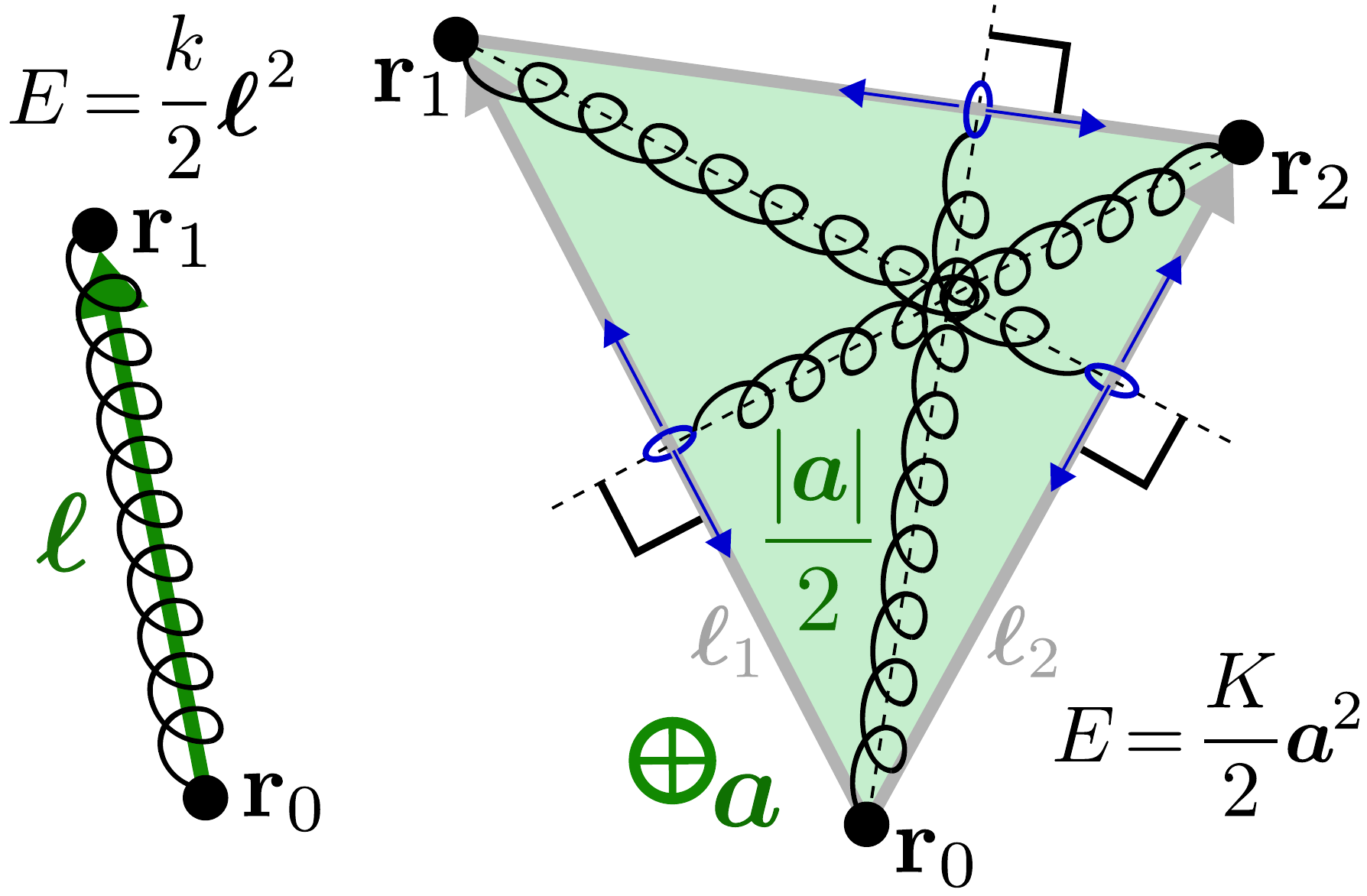} 
\caption{Left~(dumbbell model): Two beads, separated by a length vector $\boldsymbol{\ell}$, connected by a linear spring. Right~(minimal flat elastic particle model): Three beads define a vector $\boldsymbol{a}=\boldsymbol{\ell}_1\times\boldsymbol{\ell}_2$ normal to the plane of a triangle with surface area $|\boldsymbol{a}|/2$ that mimics a flat particle. An energy $E=K\boldsymbol{a}^2/2$ is produced by three identical nonlinear springs that connect each bead along the triangle's heights to the bases as shown. Spring constants are proportional to the square of the corresponding base lengths. Thus, for triangle sides as labelled here, springs connected to beads at~$\mathbf{r}_0$, $\mathbf{r}_1$ and $\mathbf{r}_2$ have spring constants $K|\boldsymbol{\ell}_1-\boldsymbol{\ell}_2|^2$, $K|\boldsymbol{\ell}_2|^2$ and $K|\boldsymbol{\ell}_1|^2$, respectively.
\label{fig:1}}
\end{figure}
The presence of elastic particles (e.g. high molecular weight polymers), is known to profoundly change the flow properties of complex liquids \cite{BAH87,Larson99,Datta22}, such as blood, saliva, and many other biological and man-made substances. For example, the extensional flow accompanying the breakup of a liquid drop leads to the formation of long and thin threads, caused by the stretching of molecules in the flow \cite{AM01,EHS20,DS19}. To model such behavior, the equations of fluid motion are usually augmented with an extra, `polymeric' or `particle' contribution $\boldsymbol{\sigma}^{\text{p}}$ to the stress tensor, found by solving an additional constitutive equation, which couples the evolution of $\boldsymbol{\sigma}^{\text{p}}$ to the flow \cite{BAH87,MS15}.

The derivation of such an equation using a realistic molecular description of a polymer has not been achieved. Instead, most constitutive equations have their basis in phenomenological models consistent with the invariances of the system as well as with thermodynamics, allowing for adjustable parameters to account for specific properties of the system \cite{BE_book,Oe_book}. 
 In particular, Oldroyd \cite{O50} pointed out that for the description to be independent of the choice of coordinate system (known as frame invariance), the ordinary convected time derivative of the stress tensor has to be augmented with extra terms, which can only take two distinct forms: the upper-convected and lower-convected derivatives (and linear superpositions thereof). However, without reference to a specific microscopic system, the phenomenological approach does not reveal which frame-invariant derivative to use, although the rheological responses can be very different. 

To address this, highly simplified model systems (known as dumbbell models, seen on the left of Fig.~\ref{fig:1}) have become very popular. They consist of two spherical beads, connected by a (harmonic) spring; their state is thus defined by the vector $\boldsymbol{\ell}=\mathbf{r}_1-\mathbf{r}_0$ alone. If both beads are convected passively by the flow, the material time derivative $\dot{\boldsymbol{\ell}}$ (which we denote by a dot) is $\mathbf{v}(\mathbf{r}_1)-\mathbf{v}(\mathbf{r}_0)$; assuming that the flow varies on scales much larger than the particle size, we can expand $\mathbf{v}$ into a Taylor series to find
 to lowest order:
\begin{equation}
\dot{\boldsymbol{\ell}}=\boldsymbol{\ell}\cdot\nabla\mathbf{v},\label{eq:dtl}
\end{equation}
with the strain rate tensor $(\nabla\mathbf{v})_{\alpha\beta}=\partial_{\alpha}v_{\beta}$ and $\alpha,\beta=x,y,z$. Using this together with Stokes' drag law \cite{LL84a}, and adding thermal noise, the thermal average of the polymeric stress yields the Oldroyd-B model \cite{Larson99}, which has become an almost canonical description of a polymeric fluid \cite{Datta22}, and which contains the upper-convected derivative only. However, depending on the geometry of the immersed particles, the lower-convected derivative should generically come in as well~\cite{hinch21}. 

In order to improve our intuition, and to expand the lexicon of distinct rheological models with an explicit microscopic underpinning, we develop in this letter an exactly solvable particle model that only contains the lower-convected derivative as part of its constitutive equation. According to long-established intuition \cite{Larson88}, p.66, and supported by a recent penetrating analysis of ellipsoidal particles \cite{SSB23}, our model involves ideally flat geometries, that are deformed by the flow. As a convenient abstraction, we consider three nearby Stokes beads (see Fig.~\ref{fig:1}, right) and two vectors $\boldsymbol{\ell}_1=\mathbf{r}_1-\mathbf{r}_0$ and $\boldsymbol{\ell}_2=\mathbf{r}_2-\mathbf{r}_0$ that uniquely define an ``area vector''
\begin{equation}\label{eq:adef}
\boldsymbol{a} = \boldsymbol{\ell}_1\times\boldsymbol{\ell}_2,
\end{equation}
which is normal to the plane of the beads, and whose modulus is twice the area of the triangle defined by the beads. 

If each of the vectors $\boldsymbol{\ell}_1$ and $\boldsymbol{\ell}_2$ evolves according to~\eqref{eq:dtl}, we find that
$\dot{\boldsymbol{a}}=\dot{\boldsymbol{\ell}}_1\times\boldsymbol{\ell}_2
+\boldsymbol{\ell}_1\times\dot{\boldsymbol{\ell}}_2=
\boldsymbol{\ell}_1\cdot\nabla\left(\mathbf{v}\times\boldsymbol{\ell}_2\right)+\boldsymbol{\ell}_2\cdot\nabla\left(\boldsymbol{\ell}_1\times\mathbf{v}\right)$. Using the vector identity \cite{suppPRL}
\begin{equation}\label{eq:cpid}
\boldsymbol{\ell}_1\cdot\nabla(\mathbf{v}\times\boldsymbol{\ell}_2)+\boldsymbol{\ell}_2\cdot\nabla(\boldsymbol{\ell}_1\times\mathbf{v})=-(\nabla\mathbf{v})\cdot(\boldsymbol{\ell}_1\times\boldsymbol{\ell}_2),
\end{equation}
which holds for $\nabla\cdot\mathbf{v}=0$, it follows that in an incompressible flow the area vector obeys
\begin{equation}\label{eq:dta}
\dot{\boldsymbol{a}}=-\boldsymbol{a}\cdot(\nabla\mathbf{v})^\top,
\end{equation}
instead of \eqref{eq:dtl} for a length vector, as expected from the dynamics of a surface area element~\cite{stone17}.

We now calculate the contribution  $\boldsymbol{\sigma}^{\text{p}}$ of a flat elastic particle suspension to the stress tensor, which is usually done in two steps
\cite{Larson99,SPHE20}: first, one derives the equation of motion for a ``fabric tensor", which characterizes the state of the suspended particle. Second, the stress is calculated from the fabric tensor by a constitutive relation. As shown in \cite{SSB23}, the equation of motion for a fabric tensor $\bsfL=c_{\ell}\langle\boldsymbol{\ell}\otimes\boldsymbol{\ell}\rangle$, appropriate for a linear dumbbell
(where $\langle\cdot\rangle$ denotes a thermal average), automatically involves the upper-convected derivative. On the other hand, in deriving the equation of motion for a fabric tensor $\bsfA=c_a\langle\boldsymbol{a}\otimes\boldsymbol{a}\rangle$ (constants $c_{\ell}$ and $c_a$ serve to make fabric tensor dimensionless), based on our flat particle model (see Fig.~\ref{fig:1}, right), we now show that the upper-convected derivative is replaced by the lower-convected derivative.
 
 According to Stokes' law \cite{LL84a}, the motion of a bead $\mathbf{r}_i$ relative to the fluid equals the force $\mathbf{F}_i$, divided by the drag coefficient $\zeta=6\pi\eta r_b$, where $\eta$ is the fluid viscosity, and $r_b$ the radius of a bead:
\begin{equation}\label{eq:FBgen}
\zeta\left(\dot{\mathbf{r}}_i-\mathbf{v}(\mathbf{r}_i)\right)=\mathbf{F}_i\equiv-\partial_{\mathbf{r}_i}E(\boldsymbol{a}),
\quad i=0,1,2.
\end{equation}
For simplicity, we focus on the deterministic part of the equation, and add thermal noise later. From now on, we assume a quadratic energy $E(\boldsymbol{\ell}_1,\boldsymbol{\ell}_2)=K\boldsymbol{a}^2/2$. We present a possible mechanical realization of this energy below (see Fig.~\ref{fig:1}, right). Due to the translational invariance of $E(\boldsymbol{\ell}_1,\boldsymbol{\ell}_2)$, the particle dynamics separates into the motion of its center of mass~$\mathbf{r}_c=(\mathbf{r}_0+\mathbf{r}_1+\mathbf{r}_2)/3$ (which is immaterial), and equations of motion for $\boldsymbol{\ell}_i$, where 
$\mathbf{r}_0=\mathbf{r}_c-(\boldsymbol{\ell}_1+\boldsymbol{\ell}_2)/3$, $\mathbf{r}_1=\mathbf{r}_c+(2\boldsymbol{\ell}_1-\boldsymbol{\ell}_2)/3$ and $\mathbf{r}_2=\mathbf{r}_c+(2\boldsymbol{\ell}_2-\boldsymbol{\ell}_1)/3$. Thus transforming to $\boldsymbol{\ell}_i$-derivatives and expanding 
$\mathbf{v}(\mathbf{r}_i)-\mathbf{v}(\mathbf{r}_0)\approx(\mathbf{r}_i-\mathbf{r}_0)\cdot\nabla\mathbf{v}$ as before, one finds
\begin{equation}
\dot{\boldsymbol{\ell}}_1=\boldsymbol{\ell}_1\cdot\nabla\mathbf{v}-\frac{1}{\zeta}\left(\frac{\partial E}{\partial{\boldsymbol{\ell}_2}}+2\frac{\partial E}{\partial
{\boldsymbol{\ell}_1}}\right),\label{eq:dtr1}
\end{equation}
and similarly for $\boldsymbol{\ell}_2$ via a permutation $1\leftrightarrow2$. Equation~\eqref{eq:dtr1} is a generalization of \eqref{eq:dtl}, describing the slip of beads relative to the flow, as a result of inter-bead forces. 

Using that 
\begin{equation}\label{eq:Eexpl}
E=K(\boldsymbol{\ell}_1^2\boldsymbol{\ell}_2^2-\left(\boldsymbol{\ell}_1\cdot\boldsymbol{\ell}_2\right)^2)/2,    
\end{equation}
one finds
\begin{equation}\label{eq:Epl}
\frac{\partial E}{\partial{\boldsymbol{\ell}_1}}=K\left(
\boldsymbol{\ell}_1\boldsymbol{\ell}_2^2-\boldsymbol{\ell}_2\boldsymbol{\ell}_1\cdot\boldsymbol{\ell}_2\right)
\end{equation}
(and correspondingly for $\boldsymbol{\ell}_2$), so that 
the last term in \eqref{eq:dtr1} reads:
\begin{equation}
\frac{\partial E}{\partial{\boldsymbol{\ell}_2}}+2\frac{\partial E}{\partial
{\boldsymbol{\ell}_1}}=\boldsymbol{\ell}_1(2\boldsymbol{\ell}_2^2-\boldsymbol{\ell}_1\cdot\boldsymbol{\ell}_2)+\boldsymbol{\ell}_2(\boldsymbol{\ell}_1^2-2\boldsymbol{\ell}_1\cdot\boldsymbol{\ell}_2).
\end{equation}
Then from the product rule $\dot{\boldsymbol{a}}=\dot{\boldsymbol{\ell}}_1\times\boldsymbol{\ell}_2
+\boldsymbol{\ell}_1\times\dot{\boldsymbol{\ell}}_2$, and repeating the calculation leading to \eqref{eq:dta}, we arrive at
 \begin{equation}\label{eq:dta_f}
\dot{\boldsymbol{a}} = -\boldsymbol{a}\cdot(\nabla\mathbf{v})^\top -\frac{9KR^2}{\zeta}\boldsymbol{a}.
\end{equation}
Here, \rev{$R^2=2[\boldsymbol{\ell}_1^2+\boldsymbol{\ell}_2^2-(\boldsymbol{\ell}_1\cdot\boldsymbol{\ell}_2)]/9$} denotes the mean-squared distance of the three beads to the center of mass~$\mathbf{r}_c$ and is thus a measure of the flat particle's size~\cite{suppPRL}. In order to make 
\eqref{eq:dta_f} amenable to a statistical analysis in analogy to the classical dumbbell model~\cite{Larson99}, we assume a size-dependent bead drag $\zeta=9R^2\zeta_a$ ($\zeta_a$ a constant), which makes \eqref{eq:dta_f} linear in $\boldsymbol{a}$. \rev{This assumption is not implausible if one imagines the particle's surface being represented by a network of beads, whose number is proportional to the surface area. It also conforms with the modeling of membranes employed by others~(e.g. \cite{Seifert1997}, pp.~100--110). However, the assumed drag also illustrates the ambiguities involved in the modeling of flat particles.} Adding thermal noise to the area vector dynamics~\cite{Gardiner}, we arrive~at
\begin{equation}\label{eq:dtastoch}
{\zeta_a}\left(\dot{\boldsymbol{a}} +\boldsymbol{a}\cdot(\nabla\mathbf{v})^\top\right)= -K\boldsymbol{a}+\boldsymbol{\xi},
\end{equation}
where $\boldsymbol{\xi}$ denotes fluctuations around equilibrium with $\langle\boldsymbol{\xi}\rangle=0$ and $\langle\xi_{\alpha}(t)\xi_{\beta}(t')\rangle=2 k_B T \zeta_a\delta_{\alpha\beta}\delta (t-t')$. The corresponding Fokker-Planck equation for the probability distribution $P(\boldsymbol{a},t)$ to find a particle with area vector~$\boldsymbol{a}$ at time $t$ then reads \cite{Gardiner}
\begin{equation}\label{eq:FP}
\frac{\partial P}{\partial t} + \frac{\partial}{\partial\boldsymbol{a}} \cdot \mathbf{J} = 0,    
\end{equation}
where the probability current is given by 
\begin{equation}\label{eq:vstoch}
\mathbf{J} = -\boldsymbol{a}\cdot(\nabla\mathbf{v})^\top P - \frac{K}{\zeta_a}\boldsymbol{a}P- \frac{k_BT}{\zeta_a}\partial_{\boldsymbol{a}} P.
\end{equation}
The thermal average to compute the fabric tensor is given by $\bsfA=c_a\int\boldsymbol{a}\otimes\boldsymbol{a}\,P(\boldsymbol{a},t)\,d\boldsymbol{a}$, such that the dynamics of the fabric tensor can be determined from the material derivative
\begin{equation}\label{eq:dCTdt}
\frac{1}{c_a}\frac{d A_{\alpha\beta}}{dt}=\int a_{\alpha}a_{\beta}\,\partial_tP\,d\boldsymbol{a}=\int (\delta_{\alpha\gamma}a_\beta+\delta_{\beta\gamma}a_{\alpha})J_{\gamma}\,d\boldsymbol{a},
\end{equation}
having used \eqref{eq:FP} and then integrated by parts. Inserting \eqref{eq:vstoch} into \eqref{eq:dCTdt}, the first two terms on the right of \eqref{eq:vstoch} are expressible directly in terms of $A_{\alpha\beta}$, the last term, integrating once more by parts, and using that $P$ is normalized, yields a constant. Thus the equation of motion \eqref{eq:dCTdt} for the fabric tensor, our first main result, becomes
\begin{equation}\label{eq:constA}
\overset\vartriangle{\bsfA} = - \frac{2 K}{\zeta_a} \bsfA + \frac{2 c_a k_B T}{\zeta_a} \mathbb{I} \equiv -\frac{1}{\lambda}(\bsfA-\mathbb{I}),
\end{equation}
where 
\begin{equation}\label{eq:LCD}
\overset\vartriangle{\bsfA}\equiv\frac{\partial\bsfA}{\partial t}+\mathbf{v}\cdot\nabla\bsfA+(\nabla\mathbf{v})\cdot\bsfA+\bsfA\cdot(\nabla\mathbf{v})^\top
\end{equation}
is the lower-convected derivative, and $\mathbb{I}$ the unit tensor. Here $\lambda = \zeta_a/(2K)$ is a relaxation time of a suspended particle, and $c_a = K/k_BT$ has been chosen to make $\bsfA$ relax toward $\mathbb{I}$ in equilibrium~\cite{SPHE20}. As anticipated, \eqref{eq:constA} is the exact analogue of the equation for the fabric tensor~$\bsfL$ of a dumbbell, but written with the lower-convected derivative. 

To compute the particle stress tensor $\boldsymbol{\sigma}^{\text{p}}\equiv\boldsymbol{\sigma}^a$ coming from our area-like elastic particles in the dilute limit, we use the Kirkwood formula \cite{DoiBook86}
\begin{equation}\label{eq:sa1}
\boldsymbol{\sigma}^a=-\rho_0\sum_{i=0}^2\langle\mathbf{r}_i\otimes\mathbf{F}_i\rangle = 
\rho_0\sum_{i=1}^2\left\langle\boldsymbol{\ell}_i\otimes\frac{\partial E}{\partial\boldsymbol{\ell}_i}\right\rangle,
\end{equation}
where $\rho_0$ is the number density of three-bead particles. In reducing to a sum over internal forces in the second step, we have used that the model flat particle is force-free, such that $\mathbf{F}_0=\partial_{\boldsymbol{\ell}_1}E+\partial_{\boldsymbol{\ell}_2}E$, $\mathbf{F}_1=-\partial_{\boldsymbol{\ell}_1}E$ and $\mathbf{F}_2=-\partial_{\boldsymbol{\ell}_2}E$. With $E(\boldsymbol{\ell}_1,\boldsymbol{\ell}_2)$ as given in Eq.~(\ref{eq:Eexpl}), the stress tensor becomes
\begin{equation}\label{eq:sa1_comp}
\boldsymbol{\sigma}^a=\rho_0 K\langle\boldsymbol{\ell}_1\otimes\boldsymbol{\ell}_1\,\boldsymbol{\ell}_2^2+
\boldsymbol{\ell}_2\otimes\boldsymbol{\ell}_2\,\boldsymbol{\ell}_1^2-2\boldsymbol{\ell}_1\otimes\boldsymbol{\ell}_2\,\boldsymbol{\ell}_1\cdot\boldsymbol{\ell}_2\rangle,
\end{equation}
and using the Levi-Civita tensor product identity for $\epsilon_{\alpha\beta\gamma}\epsilon_{\delta\epsilon\kappa}$ on $\bsfA = c_a\langle(\boldsymbol{\ell}_1\times\boldsymbol{\ell}_2)\otimes(\boldsymbol{\ell}_1\times\boldsymbol{\ell}_2)\rangle$
yields
\begin{equation}\label{eq:safin}
\boldsymbol{\sigma}^a=\mu\left[\mathbb{I}\,\text{Tr}\left(\bsfA\right)-\bsfA\right],
\end{equation}
our second main result; here $\mu=\rho_0 k_BT$ is an elastic constant. Crucially, $\boldsymbol{\sigma}^a$ is not simply proportional to the fabric tensor, as would be the case for Oldroyd models A or B, but contains an extra term involving the trace of~$\bsfA$. It is straightforward to show \cite{suppPRL} that the total stress in Eq.~(\ref{eq:safin}) corresponds to a particular case of Oldroyd's 8-constant model \cite{BAH87}. 

Finally, to complement model equations \eqref{eq:constA} and \eqref{eq:safin}, the velocity field satisfies the momentum balance 
\begin{equation}
\rho(\partial_t\mathbf{v}+\mathbf{v}\cdot\nabla\mathbf{v})
=-\nabla p + \nabla\cdot\boldsymbol{\sigma}^a + \eta\triangle\mathbf{v},
\end{equation}
where $\rho$ is the density, and $\eta$ the viscosity of the solvent. 

To illustrate the consequences of \eqref{eq:safin}, let us calculate~$\boldsymbol{\sigma}^a$ for the two-dimensional extensional flow 
\begin{equation}\label{ext}
u=\dot{\epsilon}x, \quad v=-\dot{\epsilon}y,
\end{equation}
which is stretching at a rate $\dot{\epsilon}$ in the $x$-direction, and compressing in the $y$-direction. 
Then the dynamics of the corresponding diagonal components of $\bsfA$ follow from Eq.~(\ref{eq:constA}) as
\begin{align}
\dot{A}_{xx}&=(-2\dot{\epsilon}-\lambda^{-1})A_{xx}+\lambda^{-1}\label{eq:dtAxx}\\
\dot{A}_{yy}&=(2\dot{\epsilon}-\lambda^{-1})A_{yy}+\lambda^{-1},\label{eq:dtAyy}
\end{align}
while for the fabric tensor $\bsfL$ of dumbbell particles, whose dynamics is written with an upper-convected derivative, the signs  of strain rate contributions are reversed:
\begin{align}
\dot{L}_{xx}&=(2\dot{\epsilon}-\lambda^{-1})L_{xx}+\lambda^{-1}\\
\dot{L}_{yy}&=(-2\dot{\epsilon}-\lambda^{-1})L_{yy}+\lambda^{-1}.
\end{align}
Thus for $\dot{\epsilon}>\lambda^{-1}/2$, $L_{xx}$ grows exponentially following the stretching of the flow, while the lower-convected derivative implies that $A_{yy}$ grows, in the direction in which the flow is contracting \cite{EHS20}. 

In the Oldroyd-A model, the particle stress is given by $\boldsymbol{\sigma}^p=-\mu(\bsfA-\mathbb{I})$, while for the Oldroyd-B model, \hbox{$\boldsymbol{\sigma}^p=\mu(\bsfL-\mathbb{I})$} (and stress from solvent viscosity is to be added). This means the Oldroyd-B model faithfully describes the build-up of stress in the spring of a dumbbell, as the two beads are convected by the flow (cf. Fig.~\ref{fig:1}, left). The Oldroyd-A model, on the other hand, describes the rather less intuitive situation of stress building up in a direction orthogonal to the stretching. This prediction of the Oldroyd-A model is however in stark contrast to our present finding for the behavior of idealized flat particles: Because the stress $\boldsymbol{\sigma}^a$ in Eq.~(\ref{eq:safin}) is not simply proportional to $\bsfA$, but contains an additional trace contribution, the stress in the {\it extensional} direction follows
\begin{equation}
\sigma^a_{xx}=\mu\left(\text{Tr}\left(\bsfA\right)-A_{xx}\right)\approx A_{yy} \propto A_{yy}^0 e^{(2\dot{\epsilon}-\lambda^{-1})t},
\end{equation}
and once more grows exponentially. Thus, surprisingly, the behavior of our flat elastic particle model is similar to that of a classical dumbbell model, even though the fabric tensor dynamics is given in terms of a lower-convected derivative. The two models, however, remain distinct. In an axisymmetric extensional flow, as it arises during the pinch-off of a polymeric thread \cite{AM01,EHS20}, the buildup of axial stress is weaker in our flat particle model then in the dumbbell model. As a result, no uniform thread is formed, as is observed during the pinching of solutions of elastic polymers \cite{BER90,AM01,EHS20}. \rev{On the other hand, the response of our model to a simple shear flow of rate $\dot{\gamma}$ is identical to that of the Oldroyd-A model: the first and second normal stress differences are $N_1=-N_2=2\mu\lambda^2\dot{\gamma}^2$~\cite{BAH87}. }

In conclusion, we have constructed an exactly solvable microscopic model for a dilute suspension of flat elastic particles, whose coarse-grained dynamics is described by the lower-convected derivative. Nevertheless, the rheological properties of this suspension are remarkably similar to those of conventional linear ``dumbbell" particles. 

\rev{In fact, even though this is assumed widely~\cite{hinch21}, it is very difficult to realize a flat particle model that would give rise to the Oldroyd-A model macroscopically, and in which stress builds up orthogonally to the direction of an extensional flow. The reason is that elastic forces in an overall force- and torque-free flat particle must be acting within the plane of that particle. This precludes the possibility for stresses to be generated in the direction normal to the particle's surface, which would be required to achieve the Oldroyd-A-like rheological properties described above.}

A mechanical realisation of the flat elastic particles introduced in this model is depicted in Fig.~\ref{fig:1} (right): three identical nonlinear springs with vanishing rest length connect beads along the triangle's heights to each base and have spring constants proportional to the square of the corresponding base lengths. For example, the force $\mathbf{F}_1=-\partial_{\boldsymbol{\ell}_1}E$ acting on the bead at $\mathbf{r}_1$ can according to \eqref{eq:Epl} be written as
\begin{equation}\label{eq:F1spr}
\mathbf{F}_1=-K\boldsymbol{\ell}_2^2\,\bsfP_2^{\perp}\cdot\boldsymbol{\ell}_1,   
\end{equation}
where $\bsfP_2^{\perp}=\mathbb{I}-(\boldsymbol{\ell}_2\otimes\boldsymbol{\ell}_2)/\boldsymbol{\ell}_2^2$ is a projection that extracts the component of~$\boldsymbol{\ell}_1$ that is orthogonal to $\boldsymbol{\ell}_2$. Therefore,~$\mathbf{F}_1$ is parallel to the height that originates at $\mathbf{r}_1$ and passes through the base $\boldsymbol{\ell}_2$. From \eqref{eq:F1spr}, we can identify the spring constant $K\boldsymbol{\ell}_2^2$. Similar arguments apply for the elastic forces $\mathbf{F}_2$ and $\mathbf{F}_0=-\mathbf{F}_1-\mathbf{F}_2$ acting on the beads at~$\mathbf{r}_2$ and $\mathbf{r}_0$, respectively. More general microscopic particle models can be considered, for example by using both length and area vectors as geometric building blocks. \rev{Similarly, it would be interesting to study the nonlinear dynamics and many-body effects in such a family of minimal particle models in numerical simulations.}

Finally, it would be interesting to explore the implications of our results for  rheological measurements of red blood cell suspensions~\cite{baskurt2003,nader2019,SVCOAP18}, flat objects whose mechanical properties have previously been interpreted using insights from Oldroyd-B models. \\

\paragraph{Acknowledgments:}The authors are indebted to Howard Stone for enlightening discussions and for making his work available to us prior to publication. J.E. thanks 
Antony Beris for pointing out the connection between geometry and convected derivatives.


%

\end{document}


\begin{center}
\Huge{\textbf{Supplementary Information}}
\end{center}
\begin{center}
\large{\textbf{Rheology of suspensions of flat elastic particles}} 
\\\vspace{0.2cm}


{\normalsize Jens Eggers, Tanniemola B. Liverpool, Alexander Mietke}\\\vspace{0.1cm}
{\small \textit{School of Mathematics, University of Bristol, Fry Building, Bristol BS8 1UG, United Kingdom}}
\end{center}

\normalsize

\section{I. Advective area vector dynamics}
\small
\noindent In this section, we derive the advective dynamics $\dot{\boldsymbol{a}}$ of an area vector $\boldsymbol{a}=\boldsymbol{\ell}_1\times\boldsymbol{\ell}_2$ using elementary vector calculus. For incompressible flow, $\dot{\boldsymbol{a}}=-(\nabla\mathbf{v})\cdot\boldsymbol{a}$ as stated in Eq.~(4) in the main text. In general, we have for two length vectors
\begin{align*}
\dot{\boldsymbol{\ell}}_1=\boldsymbol{\ell}_1\cdot\nabla\mathbf{v}\quad\text{and}\quad\dot{\boldsymbol{\ell}}_2=\boldsymbol{\ell}_2\cdot\nabla\mathbf{v}
\end{align*}
by the product rule the implied area vector dynamics
\begin{align}
\dot{\boldsymbol{a}}&=\boldsymbol{\ell}_1\cdot\nabla\left(\mathbf{v}\times\boldsymbol{\ell}_2\right)+\boldsymbol{\ell}_2\cdot\nabla\left(\boldsymbol{\ell}_1\times\mathbf{v}\right)\label{eq:dta1}.
\end{align}
Making use of the identity (proven below)
\begin{align}
\boldsymbol{\ell}_1\cdot\nabla(\mathbf{v}\times\boldsymbol{\ell}_2)+\boldsymbol{\ell}_2\cdot\nabla(\boldsymbol{\ell}_1\times\mathbf{v})&=(\boldsymbol{\ell}_1\times\boldsymbol{\ell}_2)\nabla\cdot\mathbf{v}-\nabla\mathbf{v}\cdot(\boldsymbol{\ell}_1\times\boldsymbol{\ell}_2),\label{eq:CPid}
\end{align}
Eq.~(\ref{eq:dta1}) becomes
\begin{align*}
\dot{\boldsymbol{a}}&=\boldsymbol{a}\,\nabla\cdot\mathbf{v}-(\nabla\mathbf{v})\cdot\boldsymbol{a}.
\end{align*}
For incompressible flow, $\nabla\cdot\mathbf{v}=0$, Eq.~(4) from the main text follows.\\

\noindent \textbf{Proof of Equation~(\ref{eq:CPid}): } We start with the known Levi-Civita tensor product identity
\begin{align}
\epsilon_{\alpha\beta\gamma}\epsilon_{\delta\epsilon\kappa}&=\delta_{\alpha\delta}\delta_{\beta\epsilon}\delta_{\gamma\kappa}+\delta_{\alpha\epsilon}\delta_{\beta\kappa}\delta_{\gamma\delta}+\delta_{\alpha\kappa}\delta_{\beta\delta}\delta_{\gamma\epsilon}-\delta_{\alpha\epsilon}\delta_{\beta\delta}\delta_{\gamma\kappa}-\delta_{\alpha\delta}\delta_{\beta\kappa}\delta_{\gamma\epsilon}-
\delta_{\alpha\kappa}\delta_{\beta\epsilon}\delta_{\gamma\delta}\nonumber\\
&=\delta_{\alpha\delta}(\delta_{\beta\epsilon}\delta_{\gamma\kappa}-\delta_{\beta\kappa}\delta_{\gamma\epsilon})+\delta_{\beta\delta}(\delta_{\gamma\epsilon}\delta_{\alpha\kappa}-\delta_{\gamma\kappa}\delta_{\alpha\epsilon})+\delta_{\gamma\delta}(\delta_{\alpha\epsilon}\delta_{\beta\kappa}-\delta_{\alpha\kappa}\delta_{\beta\epsilon})\label{eq:epseps}.
\end{align}
In the second line, we have in anticipation of the next step regrouped terms such that each term in brackets is anti-symmetric in $(\epsilon\kappa)$. Using this, as well as the standard Levi-Civita tensor property $\epsilon_{\delta\epsilon\kappa}\epsilon_{\epsilon\kappa\lambda}=2\delta_{\delta\lambda}$, the contraction of both sides of Eq.~(\ref{eq:epseps}) with $\epsilon_{\epsilon\kappa\lambda}$ over the index pair~$(\epsilon\kappa)$ yields
\begin{align}
2\delta_{\lambda\delta}\epsilon_{\alpha\beta\gamma}=2\delta_{\alpha\delta}\epsilon_{\beta\gamma\lambda}+2\delta_{\beta\delta}\epsilon_{\gamma\alpha\lambda}+2\delta_{\gamma\delta}\epsilon_{\alpha\beta\lambda}\nonumber\\
\Leftrightarrow\delta_{\lambda\delta}\epsilon_{\alpha\beta\gamma}-\delta_{\alpha\delta}\epsilon_{\lambda\beta\gamma}=\delta_{\beta\delta}\epsilon_{\alpha\lambda\gamma}+\delta_{\gamma\delta}\epsilon_{\alpha\beta\lambda}.\label{eq:ideps}
\end{align}
Contracting both sides of Eq.~(\ref{eq:ideps}) with $\partial_{\delta}v_{\lambda}(\boldsymbol{\ell}_1)_{\beta}(\boldsymbol{\ell}_2)_{\gamma}$ yields the $\alpha$-component of the desired identity in Eq.~(\ref{eq:CPid}):
\begin{align*}
&\partial_{\delta}v_{\delta}\epsilon_{\alpha\beta\gamma}(\boldsymbol{\ell}_1)_{\beta}(\boldsymbol{\ell}_2)_{\gamma}-\partial_{\alpha}v_{\lambda}\epsilon_{\lambda\beta\gamma}(\boldsymbol{\ell}_1)_{\beta}(\boldsymbol{\ell}_2)_{\gamma}=(\boldsymbol{\ell}_1)_{\delta}\partial_{\delta}\epsilon_{\alpha\lambda\gamma}v_{\lambda}(\boldsymbol{\ell}_2)_{\gamma}+(\boldsymbol{\ell}_2)_{\delta}\partial_{\delta}\epsilon_{\alpha\beta\lambda}(\boldsymbol{\ell}_1)_{\beta}v_{\lambda}\\
\Leftrightarrow\ &\nabla\cdot\mathbf{v}\,(\boldsymbol{\ell}_1\times\boldsymbol{\ell}_2)-\nabla\mathbf{v}\cdot(\boldsymbol{\ell}_1\times\boldsymbol{\ell}_2)=\boldsymbol{\ell}_1\cdot\nabla(\mathbf{v}\times\boldsymbol{\ell}_2)+\boldsymbol{\ell}_2\cdot\nabla(\boldsymbol{\ell}_1\times\mathbf{v}).
\end{align*}

\section{II. Mean-squared distance between beads and their center of mass}
\noindent In the main text, we have used that 
\begin{equation}\label{eq:Rsqsi}
R^2=\frac{2}{9}\left[\boldsymbol{\ell}_1^2+\boldsymbol{\ell}_2^2-(\boldsymbol{\ell}_1\cdot\boldsymbol{\ell}_2)\right]  
\end{equation}
corresponds to the mean-squared distance between beads and their center of mass
\begin{equation}
\mathbf{r}_c=\frac{1}{3}\left(\mathbf{r}_0+\mathbf{r}_1+\mathbf{r}_2\right).
\end{equation}
This follows directly from the definitions. To see that, recall that all bead positions can be expressed in terms of center of mass $\mathbf{r}_c$ and the length vectors $\boldsymbol{\ell}_1$ and $\boldsymbol{\ell}_2$, specifically:
\begin{align}
\mathbf{r}_0&=\mathbf{r}_c-\frac{1}{3}(\boldsymbol{\ell}_1+\boldsymbol{\ell}_2)\\
\mathbf{r}_1&=\mathbf{r}_c+\frac{1}{3}(2\boldsymbol{\ell}_1-\boldsymbol{\ell}_2)\\
\mathbf{r}_2&=\mathbf{r}_c+\frac{1}{3}(2\boldsymbol{\ell}_2-\boldsymbol{\ell}_1).    
\end{align}

From this, we can get the mean-squared distance of the beads from the centre
\begin{equation}
R^2:=\frac{1}{3}\left[\left(\mathbf{r}_0-\mathbf{r}_c\right)^2+\left(\mathbf{r}_1-\mathbf{r}_c\right)^2+\left(\mathbf{r}_2-\mathbf{r}_c\right)^2\right]=\frac{2}{9}\left[\boldsymbol{\ell}_1^2+\boldsymbol{\ell}_2^2-(\boldsymbol{\ell}_1\cdot\boldsymbol{\ell}_2)\right],    
\end{equation}
in agreement with Eq.~(\ref{eq:Rsqsi}).

\section{III. Relation to the 8-constant Oldroyd model}
\noindent The equation of motion for the fabric tensor and the constitutive equation for the particle stress~$\boldsymbol{\sigma}^a$ were found to be 
\begin{equation}\label{eq:constA}
\overset\vartriangle{\bsfA} = -\frac{1}{\lambda}(\bsfA-\mathbb{I}),
\end{equation}
and 
\begin{equation}\label{eq:safin}
\boldsymbol{\sigma}^a=\mu\left[\mathbb{I}\,\text{Tr}\left(\bsfA\right)-\bsfA\right],
\end{equation}
respectively. Both equations can be condensed into a single equation for $\boldsymbol{\sigma}^a$, by noting that 
\[
\text{Tr}\left(\overset\vartriangle{\bsfA}\right)=\text{Tr}\left(\frac{d\bsfA}{dt}\right)+\bsfA:\dot{\boldsymbol{\gamma}}
\]
and $\overset\vartriangle{\mathbb{I}}=\dot{\boldsymbol{\gamma}}$, where $\dot{\boldsymbol{\gamma}}=\nabla\mathbf{v}+(\nabla\mathbf{v})^{\top}$ is the symmetric strain rate tensor, so that 
\[
\overset\vartriangle{\boldsymbol{\sigma}^a}=\mu\left[-\frac{\text{Tr}(\bsfA)}
{\lambda} \mathbb{I}+\frac{2}{\lambda}\mathbb{I}-\bsfA:\dot{\boldsymbol{\gamma}}\mathbb{I}+\text{Tr}(\bsfA)\dot{\boldsymbol{\gamma}}+\frac{\bsfA}{\lambda}\right].
\]
Eliminating $\bsfA$ in favor of $\boldsymbol{\sigma}^a$, noting that $\text{Tr}(\boldsymbol{\sigma}^a)=2\mu\text{Tr}(\bsfA)$ and $\mathbb{I}:\dot{\boldsymbol{\gamma}}=\text{Tr}(\dot{\boldsymbol{\gamma}})=0$ for an incompressible fluid, we obtain
\[
\boldsymbol{\sigma}^a+\lambda\overset\vartriangle{\boldsymbol{\sigma}^a}-\lambda\boldsymbol{\sigma}^a:\dot{\boldsymbol{\gamma}}\mathbb{I}-\frac{\lambda}{2}\text{Tr}(\boldsymbol{\sigma}^a)\dot{\boldsymbol{\gamma}}=2\mu\mathbb{I}.
\]
To eliminate the right hand side, we put $\boldsymbol{\sigma}^a=\boldsymbol{\sigma}+2\mu\mathbb{I}$, to obtain finally
\begin{equation}\label{eq:O8}
\boldsymbol{\sigma}+\lambda\overset\triangledown{\boldsymbol{\sigma}}+\lambda\left(\boldsymbol{\sigma}\cdot\dot{\boldsymbol{\gamma}}+\dot{\boldsymbol{\gamma}}\cdot\boldsymbol{\sigma}\right)-\lambda\boldsymbol{\sigma}:\dot{\boldsymbol{\gamma}}\mathbb{I}-\frac{\lambda}{2}\text{Tr}(\boldsymbol{\sigma})\dot{\boldsymbol{\gamma}}=\lambda\mu\dot{\boldsymbol{\gamma}};
\end{equation}
here $\boldsymbol{\sigma}^a$ and $\boldsymbol{\sigma}$ only differ by a constant, which is inconsequential. In bringing the equation into the standard form \eqref{eq:O8}, we have used $\overset\vartriangle{\boldsymbol{\sigma}}=\overset\triangledown{\boldsymbol{\sigma}}+\boldsymbol{\sigma}\cdot\dot{\boldsymbol{\gamma}}+\dot{\boldsymbol{\gamma}}\cdot\boldsymbol{\sigma}$.\\ 

\noindent Comparing to \cite{BAH87}, vol 1, p.~352, we identify $\lambda_2=\lambda$, $\lambda_3=-\lambda/2$, $\lambda_4=-\lambda$, and $\eta_0=\lambda\mu$, and remembering that $\boldsymbol{\tau}\equiv-\boldsymbol{\sigma}$ in the notation of \cite{BAH87}.
If the solvent viscosity is to be included in $\boldsymbol{\sigma}$, all~8 constants are required to assume non-trivial values. 

%